\documentclass[aps,twocolumn,superscriptaddress,%
preprintnumbers,showpacs]{revtex4}
\usepackage{amsfonts,amsmath,amssymb}
\usepackage{bm,dcolumn,multirow}
\usepackage{graphicx}

\newcommand{\feyn}[1]{
  \setbox0=\hbox{\ensuremath{#1}}
  \hbox to\wd0{\hbox to0pt{\hbox to\wd0{\hss/\hss}\hss}\box0}}
\newcommand{\aB}{a_\mathrm{b}}
\newcommand{\aBF}{a_\mathrm{bf}}
\newcommand{\gB}{g_\mathrm{b}}
\newcommand{\gBF}{g_\mathrm{bf}}
\newcommand{\muB}{\mu_\mathrm{b}}
\newcommand{\Dmu}{H}
\newcommand{\nB}{n_\mathrm{b}}
\newcommand{\lv}{w}
\newcommand{\LF}{\mathcal L_\mathrm{F}}
\newcommand{\LB}{\mathcal L_\mathrm{B}}
\newcommand{\Leff}{\mathcal L_\mathrm{eff}}
\newcommand{\Lgap}{\mathcal L_\mathrm{gap}}

\begin{document}

\preprint{INT-PUB 05-31, TKYNT-05-31}
\title{Effective field theory of boson-fermion mixtures\\
       and bound fermion states on a vortex of boson superfluid}

\author{Yusuke~Nishida}
\affiliation{Department of Physics, University of Tokyo,
             Tokyo 113-0033, Japan}
\affiliation{Institute for Nuclear Theory, University of Washington, 
             Seattle, Washington 98195-1550, USA}
\author{Dam~Thanh~Son}
\affiliation{Institute for Nuclear Theory, University of Washington, 
             Seattle, Washington 98195-1550, USA}

\date{\today}

\begin{abstract} 
 We construct a Galilean invariant low-energy effective field theory of
 boson-fermion mixtures and study bound fermion states on a vortex of
 boson superfluid.  We derive
 a simple criterion to determine for which values of the fermion 
 angular momentum $l$ there exist an infinite number of bound energy
 levels. 
 We apply our formalism to two boson-fermion mixed
 systems: the dilute solution of $^3$He in $^4$He superfluid and 
 the cold polarized Fermi gas on the BEC side of the ``splitting point.''
 For the $^3$He-$^4$He mixture, we determine parameters of the
 effective theory from experimental data as functions of pressure. 
 We predict that infinitely many bound $^3$He states on a superfluid vortex
 with $l=-2,-1,0$ are realized
 in a whole range of pressure $0$--$20$ atm, where experimental data
 are available.
 As for the cold polarized Fermi gas, while only $S$-wave $(l=0)$ and
 $P$-wave $(l=\pm1)$ bound fermion states are possible in the BEC limit,
 those with higher negative angular momentum become available as one
 moves away from the BEC limit. 
\end{abstract}

\maketitle

\section{Introduction}
Quantum degenerate mixtures of bosons and fermions provide interesting
playgrounds where effects of quantum statistics become explicit.
Several experiments on the boson-fermion mixtures have been performed,
first on the classic example of the $^3$He-$^4$He
mixture~\cite{ebner70,wilks87} and more recently in cold atomic gases
such as $^6$Li-$^7$Li~\cite{truscott01,schreck01},
$^6$Li-$^{23}$Na~\cite{hadzibabic02},
$^{40}$K-$^{87}$Rb~\cite{modugno02}, and metastable triplet
$^3$He$^*$-$^4$He$^*$ mixtures~\cite{stas04}.  Molecular condensates,
formed through the BCS-BEC crossover, provide another possibility for
the boson-fermion
mixtures~\cite{regal04,bartenstein04,zwierlein04,kinast04,bourdel04}.
Stimulated by experiments, theoretical studies of the boson-fermion
mixtures predict new phenomena such as phase separation between Fermi
gas and Bose-Einstein condensates~\cite{molmer98} or sympathetic cooling
of Fermi gas by the superfluidity of Bose-Einstein
condensates~\cite{timmermans98}. 

The main goal of this paper is to investigate the bound states of
fermions on a superfluid vortex.  
Superfluid vortices were realized by the recent experiment in cold Fermi
gas throughout the BCS and BEC regimes~\cite{zeierlein05}.
The problem has been considered in the 
literature~\cite{bulgacyu03,machida05,sensarma05}, with most of the 
treatments following the original approach by Caroli, de~Gennes, and
Matricon~\cite{caroli64} (or the more rigorous approach based on the
Bogoliubov--de~Gennes equation~\cite{gygi91}). 
The reason for us to revisit the problem is that the
Caroli--de~Gennes--Matricon formalism is expected to work well only in
the BCS regime where there exist well-defined quasiparticles. 
Its application to the BEC regime where two fermions form a bound
molecule would require an exceedingly large number of mean-field
quasiparticle states for an accurate description of the strongly bound
molecules. 
It also cannot be applied to the important case of the $^3$He-$^4$He
mixture. 

We are interested in the case when the minimum of the fermion
dispersion curve $\varepsilon(\bm p)$ sits at zero momentum:
\begin{align}
 \varepsilon(\bm p)=\varepsilon_0+\frac{p^2}{2m^*}+\cdots,
 \label{eq:dispersion}
\end{align}
where $m^*$ is positive and quartic and higher order corrections are 
negligible.  
This case is opposite to the situation in the BCS theory, where the
minimum is near the Fermi momentum, but it holds for the solution of
$^3$He in $^4$He and also for cold atom gases sufficiently deep in the
BEC regime. Our technique relies on the use of an effective field
theory. The effective field theory formalism is especially powerful in
strongly coupled systems where perturbative methods are not
applicable. In our cases the details of the interactions become
unimportant, as they are encoded into a few low-energy parameters. Thus
the theory developed in this paper is universal and can be applied to
many different physical systems.

In writing down the effective field theory, we put special emphasis on
the Galilean invariance~\cite{greiter89,wingate05,andersen02}.  
The Galilean invariance as well as the global symmetries of the system
considerably restrict the possible form of the effective Lagrangian.  
We show that coupling constants appearing in the effective field theory
can be expressed through the effective fermion mass and the derivative
of fermion energy gap with the chemical potential.  The latter is 
directly related to the Bardeen-Baym-Pines (BBP) parameter.  Both
the effective fermion mass and the BBP parameter have been measured by
experiments for the $^3$He-$^4$He mixture.

After the construction of the effective field theory, we use it to the
study of bound fermion states on a vortex of the boson superfluid.  We
derive a simple criterion which tells us at which values of the angular
momentum $l$ there are an infinite number of bound energy levels.  We
will apply our effective field theory to two boson-fermion mixed
systems, a dilute solution of $^3$He in $^4$He superfluid in
Sec.~\ref{sec:He} and the BEC regime of the cold polarized Fermi gas in
Sec.~\ref{sec:Fermi_gas}, in order to predict possible angular momenta
for bound fermion states.

\section{Dilute fermions in boson superfluid \label{sec:He}}

\subsection{Galilean invariance and effective field theory
  \label{sec:EFT}}
Here we demonstrate how to construct an effective field theory of
dilute fermions in a boson superfluid in accordance with the Galilean
invariance and global symmetries. 
Since the microscopic theory is Galilean invariant, its low-energy
effective Lagrangian should satisfy the following identity required by
the Galilean invariance:
\begin{align}
 T_{0i}=mJ_i\,, 
\end{align}
where $T_{0i}$ is the momentum density, $m$ is the mass of the particle, 
and $J_i$ is the particle number current \cite{greiter89,wingate05}. 
If there are more than one species of particles in the system, the right 
hand side of the equation is a sum over all species. 
This identity simply indicates that the momentum density should be equal
to the mass carried by the particle number current in the Galilean
invariant system.
The Galilean invariance as well as the global symmetries of the system
considerably restrict the possible form of the effective Lagrangian.
For definiteness, we consider the particular case  
of the dilute solution of
$^3\mathrm{He}$ in the $^4\mathrm{He}$ superfluid from now on.
However, the universality of the effective field theory means that our
discussion, with only minimal modifications, 
will be applicable to any other dilute fermion excitations in a boson
superfluid.

\subsubsection{$^4$He superfluid phonon}
First of all, let us consider an effective field theory for the pure
$^4$He superfluid. 
The physical degree of freedom at long distance is the phase of the
condensate $\vartheta$, which is defined as
$\langle\Psi\rangle=|\langle\Psi\rangle|e^{i\vartheta}$ with $\Psi$
being a field for the $^4$He atoms. 
Under the U(1) symmetry associated with the number conservation of
$^4$He atoms, the field $\vartheta$ transforms as
$\vartheta\to\vartheta+\chi$. 
Therefore the effective Lagrangian of the $^4$He superfluid obeying this 
symmetry should be a function of derivatives of the field
$\dot\vartheta$ and $\partial_i\vartheta$. 
We note that terms where derivatives act more than once on one field
such as $\ddot\vartheta$ or $\bm\partial^2\vartheta$ are also possible. 
However, they are in higher orders in the power counting scheme where
$\dot\vartheta$ and $\partial_i\vartheta$ are regarded as order $O(1)$. 
In consideration of the rotational symmetry, the leading order
effective Lagrangian $\mathcal L_4$ should be written by a polynomial of
$\dot\vartheta$ and $(\bm\partial\vartheta)^2$ as follows: 
\begin{align}
 \mathcal L_4=\mathcal L_4(\dot\vartheta,(\bm\partial\vartheta)^2)\,.
\end{align}

Now we impose the Galilean invariance $T_{0i}=m_4J^{(4)}_i$ on the
effective Lagrangian, where $m_4$ and $J^{(4)}_i$ are the mass and 
the number current of the $^4$He atoms. 
Since the momentum density and the particle number current are given by 
\begin{align}
 T_{0i}&=\frac{\delta\mathcal L_4}{\delta\dot\vartheta}
 \partial_i\vartheta\,,\\
 J^{(4)}_i&=\frac{\delta\mathcal L_4}{\delta\partial_i\vartheta}
 =2\partial_i\vartheta
 \frac{\delta\mathcal L_4}{\delta(\bm\partial\vartheta)^2}\,,
\end{align}
the Galilean invariance $T_{0i}=m_4J^{(4)}_i$ results in
\begin{align}
 \frac{\delta\mathcal L_4}{\delta\dot\vartheta}
 =2m_4\frac{\delta\mathcal L_4}{\delta(\bm\partial\vartheta)^2}\,.
\end{align}
This equality requires that the effective Lagrangian depends on the
field $\vartheta$ only through the combination
$\dot\vartheta+(\bm\partial\vartheta)^2/2m_4$ as follows:
\begin{align}
 \mathcal L_4=P\left(-\dot\vartheta
 -\frac{(\bm\partial\vartheta)^2}{2m_4}\right)\,. \label{eq:L4_theta}
\end{align}
Here $P(\cdots)$ is an arbitrary polynomial, which will be identified 
to the pressure of the pure $^4$He superfluid as a function of the
chemical potential later. 

In the superfluid phase, the symmetry $\vartheta\to\vartheta+\chi$ is
spontaneously broken. At finite chemical potential $\mu_4$, the ground
state of the superfluid system corresponds to $\vartheta_0=-\mu_4 t$
\cite{wingate05}. 
Then we expand the field around the ground state as
$\vartheta=\vartheta_0+\varphi$, where fluctuations of $\varphi$ around
zero corresponds to superfluid phonon excitations. 
Substitution of the expression $\vartheta=-\mu_4 t+\varphi$ into
Eq.~(\ref{eq:L4_theta}) results in 
\begin{align}
 \mathcal L_4=P\left(\mu_4-\dot\varphi
 -\frac{(\bm\partial\varphi)^2}{2m_4}\right)\,. \label{eq:L4_phi}
\end{align}
Now we can show that the function $P(\cdots)$ is identical to the
pressure as a function of $\mu_4$ at zero temperature up to an irrelevant 
constant. 
For that purpose, we calculate the number density of $^4$He atoms $n_4$
by differentiate the Lagrangian with $\mu_4$ and we obtain
\begin{align}
 n_4(\mu_4)=\frac{\partial\mathcal L_4}{\partial\mu_4}
 =\frac{\partial P}{\partial\mu_4}\,.
\end{align}
This equation implies that $P$ is a function of the chemical potential
satisfying $P'(\mu_4)=n_4(\mu_4)$, which means that $P(\mu_4)$ is
identical to the thermodynamic pressure up to an irrelevant constant. 
Once equation of state of the $^4$He superfluid $P(\mu_4)$ is given as a 
function of the chemical potential, the low-energy effective field
theory of superfluid phonons is simply given by replacing $\mu_4$ with 
$\mu_4-\dot\varphi-(\bm\partial\varphi)^2/2m_4$.

In order to proceed our analysis further, we expand
Eq.~(\ref{eq:L4_phi}) up to the second order in fields
\begin{align}
 \mathcal L_4&\simeq P(\mu_4)-\frac{\partial P}{\partial\mu_4}
 \left[\dot\varphi+\frac{(\bm\partial\varphi)^2}{2m_4}\right]
 +\frac12\frac{\partial^2P}{\partial\mu_4^2}
 \left[\dot\varphi+\frac{(\bm\partial\varphi)^2}{2m_4}\right]^2
 \nonumber\\
 &\simeq P(\mu_4)-n_4\dot\varphi
 +\frac{\partial n_4}{\partial\mu_4}\frac{\dot\varphi^2}2
 -\frac{n_4}{m_4}\frac{(\bm\partial\varphi)^2}2\,.
 \label{eq:L4}
\end{align}
The first term gives the pressure without phonon excitations.
The second term is a total derivative of the field, which does not
affect the equation of motion. 
The third and fourth terms represent the propagation of phonon with its
sound velocity \cite{wingate05,son02}
\begin{align}
 \sqrt{\frac{n_4}{m_4}\frac{\partial\mu_4}{\partial n_4}}
 =\sqrt{\frac{\partial P}{m_4\,\partial n_4}}\,.
\end{align}
The higher order terms which are not shown in Eq.~(\ref{eq:L4})
represent self-interactions among phonons. 

\subsubsection{Minimal coupling between $^4$He and $^3$He}
Next, we consider the coupling between the superfluid phonon $\varphi$ and
a $^3$He atom $\psi$.  One coupling term can be written down from the
following argument.  We note that there is an energy cost to
introduce a single $^3$He atom into the $^4$He superfluid.
This energy is some function $\Delta(\mu_4)$ of the $^4$He chemical
potential $\mu_4$.  
However, Galilean invariance tells us that $\mu_4$ always enters the
Lagrangian in the combination
$\mu_4-\dot\varphi-(\bm\partial\varphi)^2/2m_4$.  
Thus, the Lagrangian contains the following term:
\begin{align}
 \Lgap&=-\Delta
 \left(\mu_4-\dot\varphi-\frac{(\bm\partial\varphi)^2}{2m_4}\right)
 \psi^\dagger\psi\nonumber\\
 &\simeq-\Delta(\mu_4)\,\psi^\dagger\psi
 +\frac{\partial\Delta}{\partial\mu_4}
 \left[\dot\varphi+\frac{(\bm\partial\varphi)^2}{2m_4}\right]
 \psi^\dagger\psi\,. \label{eq:Lgap}
\end{align}
The first term represents the energy cost of introducing a $^3$He atom
into the pure $^4$He superfluid. 
The second term proportional to $\partial\Delta/\partial\mu_4$ gives a
Galilean invariant coupling between the superfluid phonon and
the $^3$He atom put into the $^4$He superfluid.  However, this is not
the only coupling between $^3$He and $^4$He, as we shall see below.

\subsubsection{$^3$He kinetic term}
Now, let us consider the kinetic term of the $^3$He field $\psi$.
The $^3$He atom put into the $^4$He superfluid has the effective mass
$m_3^*$, not equal to the bare mass $m_3$ due to the strong
interaction with the $^4$He superfluid.  Thus the
kinetic term of the $^3$He atom is
\begin{align}
 \mathcal L_3(\psi,\psi^\dagger)
 =\frac{i\psi^\dagger\tensor\partial_0\psi}2
 -\frac{\left|\bm\partial\psi\right|^2}{2m_3^*}
 +\mu_3\psi^\dagger\psi \label{eq:L3_bare}
\end{align}
with $\mu_3$ being the chemical potential for the $^3$He atom.
However, this Lagrangian does not satisfy the Galilean invariance
$T_{0i}=m_3J^{(3)}_i$, where $J^{(3)}_i$ is the number current of the
$^3$He atom. 
This is because the Galilean invariance condition involves the bare
mass, while the kinetic term in the Lagrangian involves the effective
mass. 

The resolution to this apparent paradox is that there are interaction
terms that contribute to both the momentum density and the particle
number current in order to restore the Galilean invariance.  
Therefore, the Galilean invariance imposes some relationships between
the interaction terms and the $^3$He effective mass. 
This situation here is reminiscent of the Fermi liquid theory, where
there exists a relationship between an effective fermion mass and a
Landau parameter.

One way to construct the Lagrangian that obeys the Galilean invariance
is as follows \cite{greiter89}. 
The effective mass of $^3$He atom $m_3^*$ originates in the fact that
the $^3$He quasiparticle in the $^4$He superfluid entrains superfluid
$^4$He atoms due to the strong interaction between them.  
Therefore, an elementary excitation of the system, or the $^3$He
quasiparticle $\tilde\psi$, should be written as
$\tilde\psi=e^{i\eta\varphi}\psi$.  
Here $\eta$ is a parameter defined by $m_3^*=m_3+\eta\,m_4$, which
represents a ``fraction'' of superfluid $^4$He atoms in the $^3$He
quasiparticle.  
Using the $^3$He quasiparticle field $\tilde\psi$ instead of the bare
$^3$He field $\psi$ in Eq.~(\ref{eq:L3_bare}), we have
\begin{align}
 \mathcal L_3(\tilde\psi,\tilde\psi^\dagger)
 &=\frac{i\psi^\dagger\tensor\partial_0\psi}2
 -\frac{\left|\bm\partial\psi\right|^2}{2m_3^*}
 +\mu_3\psi^\dagger\psi \label{eq:L3}\\
 &\quad+\eta\,\bm\partial\varphi\cdot
 \frac{i\psi^\dagger\tensor{\bm\partial}\psi}{2m_3^*}
 -\biggl[\eta\,\dot\varphi
 +\eta^2\frac{\left(\bm\partial\varphi\right)^2}
 {2m_3^*}\biggr]\psi^\dagger\psi\,.\nonumber
\end{align}

We can verify that this modified Lagrangian (\ref{eq:L3}) obeys the
Galilean invariance. 
The momentum density of the system is given by
\begin{align}
 T_{0i}&=\frac{\delta\mathcal L_3}{\delta\dot{\psi}}\,\partial_i\psi
 +\frac{\delta\mathcal L_3}{\delta\dot{\psi}^\dagger}\,
 \partial_i\psi^\dagger+\frac{\delta\mathcal L_3}{\delta\dot{\varphi}}
 \,\partial_i\varphi\nonumber\\
 &=\frac{i\psi^\dagger\tensor{\partial}_i\psi}2
 -\eta\,\partial_i\varphi\,\psi^\dagger\psi\,.
\end{align}
On the other hand, the number currents associated with the
$^3$He atoms and $^4$He atoms are, respectively, given by
\begin{align}
 J^{(3)}_i&=\frac{\delta\mathcal L_3}{\delta\partial_i\psi}\,i\psi
 -\frac{\delta\mathcal L_3}{\delta\partial_i\psi^\dagger}\,i\psi^\dagger
 =\frac{i\psi^\dagger\tensor{\partial}_i\psi}{2m_3^*}
 -\eta\frac{\partial_i\varphi}{m_3^*}\psi^\dagger\psi
\intertext{and}
 J^{(4)}_i&=\frac{\delta\mathcal L_3}{\delta\partial_i\varphi}
 =\eta\frac{i\psi^\dagger\tensor{\partial}_i\psi}{2m_3^*}
 -\eta^2\frac{\partial_i\varphi}{m_3^*}\psi^\dagger\psi\,.
\end{align}
Recalling the definition of $\eta$, the Galilean invariance
$T_{0i}=m_3J^{(3)}_i+m_4J^{(4)}_i$ is indeed satisfied by the Lagrangian
(\ref{eq:L3}). 
We note that the $^3$He self-interactions are negligible in the dilute
limit of the $^3$He density, while in general they appear as higher
order terms in the Lagrangian. 

\subsubsection{Effective Lagrangian}
Finally, getting Eqs.~(\ref{eq:L4}), (\ref{eq:Lgap}) and (\ref{eq:L3})
together, we have the low-energy effective Lagrangian of the $^4$He
superfluid phonons and $^3$He excitations as follows:
\begin{equation}
\begin{split}
 \Leff&=\frac{f_t^2}2\dot{\varphi}^2
 -\frac{f^2}2\left(\bm\partial\varphi\right)^2\\
 &\quad+\frac{i\psi^\dagger\tensor\partial_0\psi}2
 -\frac{\left|\bm\partial\psi\right|^2}{2m_3^*}
 +\left(\mu_3-\Delta\right)\psi^\dagger\psi\\
 &\quad+g_1\bm\partial\varphi\cdot
 \frac{i\psi^\dagger\tensor{\bm\partial}\psi}{2m_4}
 +\biggl[g_2\dot\varphi+g_3\frac{\left(\bm\partial\varphi\right)^2}{2m_4}
 \biggr]\psi^\dagger\psi\,.\label{eq:Leff}
\end{split}
\end{equation}
Here we have introduced low-energy parameters as
\begin{align}
f_t^2=\frac{\partial n_4}{\partial\mu_4}\,,\qquad f^2=\frac{n_4}{m_4}
\end{align}
and couplings as
\begin{align}
 g_1&=\eta\frac{m_4}{m_3^*}\,,&
 g_2&=\frac{\partial\Delta}{\partial\mu_4}-\eta\,,&
 g_3&=\frac{\partial\Delta}{\partial\mu_4}-\eta^2\frac{m_4}{m_3^*}\,.
 \label{eq:g_123}
\end{align}
While there are three independent terms representing interactions
between the $^4$He superfluid phonons and $^3$He excitations, their
couplings are not independent. 
We have one constraint on the three couplings
\begin{align}
 g_3=g_2+\frac{m_3}{m_4}g_1
\end{align}
as a consequence of the Galilean invariance.  
Moreover $g_1$ is completely determined by the $^3$He effective mass
$m_3^*$, and $g_2$ (and therefore $g_3$) is determined by $m_3^*$ and
the derivative of the energy cost for introducing a $^3$He atom with the 
chemical potential $\Delta'(\mu_4)$. 
These free parameters should be either determined by experiments or 
computed from microscopic theories. 
We emphasize again that our low-energy effective Lagrangian
(\ref{eq:Leff}) and its consequences are applicable to any other dilute
fermion excitations in a boson superfluid. 
Details of strong interactions among bare particles are encoded
into the effective mass $m^*$ and the energy cost function
$\Delta(\mu)$.

\subsection{Bound states on a superfluid vortex  \label{sec:BS}}
One of the consequences derived from the effective Lagrangian
is on bound $^3$He states on a vortex of $^4$He 
superfluid.  
Let us consider a single vortex with its winding number $\lv$ in the
$^4$He superfluid.  
The phase $\varphi$ around the vortex is given by $\varphi(t,\bm
r)=\lv\theta$ in cylindrical coordinates. 

The equation of motion of the field $\psi$ from Eq.~(\ref{eq:Leff})
gives a Schr\"odinger equation for a $^3$He atom on the vortex
$\varphi=\lv\theta$ as
follows: 
\begin{align}
 \tilde E\psi(\bm r)&=-\frac{\bm\partial^2\psi(\bm r)}{2m_3^*}
 -\frac{ig_1\lv}{m_4r^2}\frac{\partial\psi(\bm r)}{\partial\theta}
 -\frac{g_3\lv^2}{2m_4r^2}\psi(\bm r)\,. 
\end{align}
$\tilde E$ is the energy eigenvalue of the $^3$He atom and is negative
for bound states.  
Separation of variables leads to the equation for the radial wave
function $R(r)$ as
\begin{align}
 &\left(2m_3^*\tilde E-k_z^2\right)R(r) \label{eq:radial_He}\\
 &\ \ =\left[-\frac{\partial^2}{\partial r^2}
 -\frac1r\frac{\partial}{\partial r}
 +\frac{l^2}{r^2}+2g_1\frac{m_3^*}{m_4}\frac{\lv l}{r^2}
 -g_3\frac{m_3^*}{m_4}\frac{\lv^2}{r^2}\right]R(r)\,. \nonumber
\end{align}
Here $l$ is an angular momentum of the $^3$He atom and $k_z$ is its
momentum along the vortex line. 
Introducing $E=\tilde E-k_z^2/2m_3^*<0$ and 
\begin{align}
 \kappa=\sqrt{g_3\frac{m_3^*}{m_4}\lv^2
 -2g_1\frac{m_3^*}{m_4}\lv l-l^2}\,,
\end{align}
we can rewrite Eq.~(\ref{eq:radial_He}) as 
\begin{align}
 \frac{\partial^2R}{\partial r^2}+\frac1r\frac{\partial R}{\partial r}
 -\left[2m_3^*\left|E\right|-\frac{\kappa^2}{r^2}\right]R=0\,.
 \label{eq:bessel}
\end{align} 
Solutions of this equation are given in terms of the modified Bessel
functions as 
$R(r)=I_{\pm i\kappa}(\sqrt{2m_3^*\left|E\right|}\,r)$.

One should remind that our effective field theory written in terms
of the phase of the condensate is valid only far away from the vortex
core where the magnitude of the condensate is almost constant. 
Suppose the Schr\"odinger equation (\ref{eq:bessel}) is valid for 
$r\geq r_0$ and the binding energy $\left|E\right|$ is 
small enough to satisfy $\sqrt{2m_3^*\left|E\right|}\,r_0\ll1$. 
In this instance, the solution of Eq.~(\ref{eq:bessel}) near the point
$r=r_0$ turns out to be
\begin{align}
 R(r)=C\sin\left[\kappa\ln\Bigl(\sqrt{2m_3^*\left|E\right|}
 \,r\Bigr)-\phi\right] \label{eq:solution}
\end{align}
with $C$ and $\phi$ being arbitrary constants. 
This radial wave function for $r\agt r_0$ should smoothly connect with
the radial wave function from $r\alt r_0$. 
If the binding energy $\left|E\right|$ is small enough compared to
the potential energy at $r\alt r_0$, the radial wave function for $r\alt
r_0$ will not depend on $\left|E\right|$. 
Therefore, the solution (\ref{eq:solution}) is required to be
independent of $\left|E\right|$ at $r=r_0$. 

The logarithmic derivative of Eq.~(\ref{eq:solution}) at $r=r_0$ results
in 
\begin{align}
 \frac{R'(r_0)}{R(r_0)}=\frac{\kappa}{r_0}\cot\left[\kappa
 \ln\Bigl(\sqrt{2m_3^*\left|E\right|}\,r_0\Bigr)-\phi\right].
\end{align}
In order to satisfy the requirement, allowed energy levels
$\left|E\right|$ should be discretized as  
\begin{align}
 \kappa\ln\Bigl(\sqrt{2m_3^*\left|E_n\right|}\,r_0\Bigr)=D-n\pi
\end{align}
or, equivalently, 
\begin{align}
 E_n=-\frac{e^{2D/\kappa}}{2m_3^*r_0^{\,2}}e^{-2n\pi/\kappa}.
\end{align}
Here $D$ is an arbitrary constant and $n$ is an arbitrary integer large
enough to satisfy $\left|E_n\right|\ll1/(2m_3^*r_0^{\,2})$.
Therefore, as long as $\kappa^2$ is positive, we have an infinite number
of energy levels for bound $^3$He states. While we can not determine the
absolute values of bound energy levels in our approach, their ratios are
independent of unknown constants $r_0$, $D$ and are asymptotically
given by
\begin{align}
 \frac{E_n}{E_{n-1}}=e^{-2\pi/\kappa}
 \qquad\quad\text{for \ }n\to\infty\,. \label{eq:ratio_He}
\end{align}

\begin{table*}[tp]
 \caption{Pressure dependence of measured quantities $m_3^*$ and
 $\alpha_0$ and calculated quantities $g_{1,2,3}$ and $l_\pm$ at zero
 temperature and zero $^3$He concentration. 
 The ratio of the $^3$He effective mass to its bare mass $m_3^*(P)/m_3$
 is taken from Ref.~\cite{krotscheck98}, which is obtained by fitting
 experimental data \cite{yorozu93} 
 to the zero concentration. 
 The BBP parameter at zero concentration $\alpha_0(P)$ is determined up
 to 10 atm by the experiment in Ref.~\cite{hatakeyama03}. 
 $\alpha_0(P)$ for 15, 20 atm (indicated with daggers) are read from
 Ref.~\cite{boghosian67}, while they are at 6.0\% concentration and not
 extrapolated to the zero concentration limit.
 The couplings $g_{1,2,3}$ in the effective Lagrangian and $l_\pm$ in 
 Eq.(\ref{eq:l+/-}) are calculated from the measured values with the use
 of Eqs.~(\ref{eq:g_123}) and (\ref{eq:Delta}).
 \label{tab:pressure}}\smallskip 
 \begin{ruledtabular}
  \begin{tabular}{cccccccc}
   $\quad\ P$ [atm]&$m_3^*(P)/m_3$&$\alpha_0(P)\phantom{^*}$
   &$g_1$&$\,g_2$&$g_3$&$\ l_-$&$l_+\qquad$\\
   \hline
   \phantom{1}0&2.18&0.288\phantom{$^\dagger$}
   &0.541&\phantom{$-$}0.403\phantom{0}&0.809&$-2.34$&0.566$\qquad$\\
   \phantom{1}5&2.31&0.242\phantom{$^\dagger$}
   &0.567&\phantom{$-$}0.260\phantom{0}&0.685&$-2.45$&0.485$\qquad$\\
   10&2.44&0.218\phantom{$^\dagger$}
   &0.590&\phantom{$-$}0.138\phantom{0}&0.581&$-2.57$&0.413$\qquad$\\
   15&2.54&0.172$^\dagger$
   &0.606&\phantom{$-$}0.0170&0.472&$-2.65$&0.338$\qquad$\\
   20&2.64&0.148$^\dagger$
   &0.621&$-0.0820$&0.384&$-2.74$&0.278$\qquad$
  \end{tabular}
 \end{ruledtabular}
\end{table*}

The criterion for an angular momentum $l$ of $^3$He atom in which bound
$^3$He states are available is simply given by
\begin{align}
 \kappa^2=g_3\frac{m_3^*}{m_4}\lv^2-2g_1\frac{m_3^*}{m_4}\lv l
 -l^2>0\,. \label{eq:criterion_He}
\end{align}
Using the definitions of the couplings (\ref{eq:g_123}) and $\eta$,
$m_3^*=m_3+\eta\,m_4$, the criterion can be rewritten as $l_-<l<l_+$
with 
\begin{align}
 \frac{l_\pm}{\lv}=-\frac{m_3^*-m_3}{m_4}
 \pm\sqrt{\frac{\partial\Delta}{\partial\mu_4}\frac{m_3^*}{m_4}}\,.
 \label{eq:l+/-}
\end{align}
An infinite number of bound energy levels with their asymptotic ratio
$e^{-2\pi/\kappa}$ appear for each integer $l$ satisfying $l_-<l<l_+$. 
We note that binding energies themselves are not determined within the
effective Lagrangian, because it can not access the vicinity of the
vortex core where the magnitude of the condensate changes.

\subsection{Determination of parameters and results for
  $^3\mathrm{He}$-$^4\mathrm{He}$ mixture}

We should determine the free parameters of our theory $m_3^*$ and
$\partial\Delta/\partial\mu_4$ from experiments. 
We adopt the $^3$He effective mass $m_3^*$ at zero temperature from
Ref.~\cite{krotscheck98}, which is obtained by fitting experimental data 
\cite{yorozu93} to the zero $^3$He concentration. 
The ratio to the bare $^3$He mass $m_3^*(P)/m_3$ is cited in
Table~\ref{tab:pressure} at various pressures $P=0,5,10,15$, and 20 atm. 

On the other hand, 
the derivative of the $^3$He energy cost with the $^4$He chemical
potential $\partial\Delta/\partial\mu_4$ is related to the relative 
fractional molar volume of $^3$He in a $^3$He-$^4$He solution, or the
Bardeen-Baym-Pines (BBP) parameter \cite{bardeen67}, at zero
concentration $\alpha_0$ by \cite{baym-pethick}
\begin{align}
 \frac{\partial\Delta}{\partial\mu_4}=\alpha_0(P)+1\,. 
 \label{eq:Delta}
\end{align}
$\alpha_0$ has been determined as a function of the pressure $P$ up to
about 10 atm by the experiment \cite{hatakeyama03}, and is shown in
Table~\ref{tab:pressure}. 
The values for 15, 20 atm in Table~\ref{tab:pressure} are quoted from
Ref.~\cite{boghosian67}, while they are at 6.0\% concentration and not
extrapolated to the zero concentration limit.

Then, $l_\pm$ in Eq.~(\ref{eq:l+/-}) as well as the couplings
$g_{1,2,3}$ in our effective Lagrangian (\ref{eq:g_123}) are determined
from experimental values as functions of the pressure, which are listed
in Table~\ref{tab:pressure}.  We have put $\lv=1$ because only $|\lv|=1$ 
vortex is energetically stable.  The results show that bound $^3$He
states on a vortex of the $^4$He superfluid are realized in $l=0$,
$l=-1$, and $l=-2$ channels in a whole range of pressure, $0$--$20$
atm, where experimental data are available.  
Note that because parity is broken by the vortex, the $l=1$ and $l=2$
states are not bound while the $l=-1$ and $l=-2$ states are bound. 
The asymptotic ratios of energy levels (\ref{eq:ratio_He}) of bound
$^3$He states for each possible angular momentum are shown at
$P=0,5,10,15$, and 20 atm in Table~\ref{tab:ratio}.
We conclude that those values are in principle measurable by accurate
experiments. 

\begin{table}[tp]
 \caption{Asymptotic ratios of energy levels
 $E_\infty/E_{\infty-1}=e^{-2\pi/\kappa}$ of bound $^3$He states for
 each possible angular momentum at pressures $P=0,5,10,15$, and 20 atm. 
 \label{tab:ratio}}\smallskip 
 \begin{ruledtabular}
  \begin{tabular}{cccc}
   $\ P$ [atm]&$l=0$&$l=-1$&$l=-2\ $\\\hline
   \phantom{1}0&$4.24\times10^{-3}$
   &$1.30\times10^{-2}$&$1.15\times10^{-3}\ $\\
   \phantom{1}5&$3.13\times10^{-3}$
   &$1.38\times10^{-2}$&$2.62\times10^{-3}\ $\\
   10&$2.25\times10^{-3}$&$1.48\times10^{-2}$&$4.48\times10^{-3}\ $\\
   15&$1.32\times10^{-3}$&$1.46\times10^{-2}$&$6.11\times10^{-3}\ $\\
   20&$7.42\times10^{-4}$&$1.47\times10^{-2}$&$7.85\times10^{-3}\ $
  \end{tabular}
 \end{ruledtabular}
\end{table}

\section{Cold Fermi gas \label{sec:Fermi_gas}}

The effective field theory described above makes no assumption about
the nature of the bosonic ($\varphi$) and fermionic ($\psi$) degrees
of freedom.  This means we can use it for a two-component Fermi gas
when $\varphi$ is the phase of the Cooper pair and $\psi$ represents
the fermion quasiparticles of a chosen component (say, the spin-up
component if the two components correspond to different spins).  The
fact that the Cooper pair is made up from the two fermions, one of
which is the $\psi$ fermion, is of no importance from the point of
view of an effective field theory.

Let us consider the cold two-component Fermi gas with a scattering length
$a$, which has been recently archived by experiments using the technique
of Feshbach resonance 
\cite{regal04,bartenstein04,zwierlein04,kinast04,bourdel04}.
Its ground state is found to be the superfluid in a whole range of $a$
via usual BCS mechanism for $a<0$ (BCS regime) 
or Bose-Einstein condensation of tightly bound Cooper pairs (molecules)
for $a>0$ (BEC regime) \cite{eagles69,legett80,nozieres85}. 

One can have in mind the situation of a slightly unequal number
densities in the two fermion components.  
Then, the ground state would be a homogeneous mixture of superfluid
Cooper pairs and extra dilute fermions carrying the single component.
Since the minimum of the fermion dispersion curve sits at zero momentum
sufficiently deep in the BEC regime (corresponding to phase III in
Ref.~\cite{son05}), such a system turns out to be within the scope of
our low-energy effective field theory for boson-fermion mixtures. 
As in the case for $^3$He-$^4$He mixture, we are interested in the
situation when the difference in the number densities is very small so
that the interactions among the extra fermions are negligible. 

In our subsequent discussion, we refer to the inequality in the number
densities as ``polarization'' according to the terminology of
Ref.~\cite{son05} in the case of spin-$\frac12$ fermions.
Also, we shall use the term ``BEC regime'' for the case $a>0$, and ``BEC
limit'' for the limit $na^3\to+0$. 
We introduce the notation $\mu=(\mu_\uparrow+\mu_\downarrow)/2$ and  
$\Dmu=(\mu_\uparrow-\mu_\downarrow)/2$, with $\mu_\uparrow$ and 
$\mu_\downarrow$ being chemical potentials of each component of fermions
for later use.

\subsection{BEC limit}

\subsubsection{Microscopic description of the system}
It would be instructive to start with the microscopic description of the 
system in the BEC limit $na^3\ll1$, where $n$ is the fermion number
density without polarization. 
Since the system is dilute in this limit, the bound molecules can be
regarded as pointlike bosons. 
Therefore, the dynamics of the molecules could be described by the
following local Lagrangian 
\cite{melo93}: 
\begin{align}
 \LB=\frac{i\Psi^\dagger\tensor\partial_0\Psi}2
 -\frac{\left|\bm\partial\Psi\right|^2}{4m}+\muB\Psi^\dagger\Psi
 -\frac\gB2\left(\Psi^\dagger\Psi\right)^2. \label{eq:boson}
\end{align}
Here the field $\Psi$ represents the superfluid molecule which has its
mass $2m$ with $m$ being the fermion mass.
$\muB=2\mu+E_0$ is the chemical potential of molecule with $E_0$ being
its binding energy. 
The coupling of its self-interaction $\gB$ is characterized in terms
of the two-body scattering length between molecules by $\gB=2\pi\aB/m$. 

If we introduce the chemical potential for the polarization $\Dmu$
larger than the energy gap of a single fermion 
$\left|\Dmu\right|\agt E_0/2$, extra fermions carrying one sign of
spin will be created on the top of the BEC ground state.
The Lagrangian describing such fermions and their interaction with the 
superfluid molecules will be given by 
\begin{align}
 \LF=\frac{i\psi^\dagger\tensor\partial_0\psi}2
 -\frac{\left|\bm\partial\psi\right|^2}{2m}
 +\left(\mu+\left|\Dmu\right|\right)\psi^\dagger\psi
 -\gBF\Psi^\dagger\Psi\,\psi^\dagger\psi\,. \label{eq:fermion}
\end{align}
$\psi=\psi_\uparrow$ or $\psi_\downarrow$ depending on the sign of
$\Dmu$. 
Hereafter $\Dmu>0$ is assumed to be positive without losing
generality. 
The coupling of the interaction between the extra fermion and the bound
molecule $\gBF$ is characterized in terms of their two-body scattering
length by $\gBF=3\pi\aBF/m$. 
Self-interactions among fermions are negligible in the dilute limit of
extra fermions. 
We should note here that the mass of the extra fermion is provided by
its bare mass $m$, because interaction effects become infinitely small
in the BEC limit $na^3\to0$. 

\begin{figure}[tp]
 \includegraphics[width=0.45\textwidth,clip]{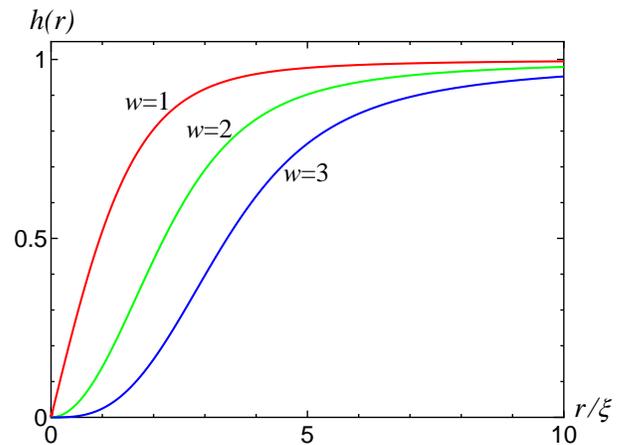}
 \caption{(Color online) Solutions of Eq.~(\ref{eq:h(x)}) as functions
 of $r/\xi$ for $\lv=1,2$, and $3$. \label{fig:GP-solution}}
\end{figure}

The equation of motion from Eq.~(\ref{eq:boson}) leads to the 
Gross-Pitaevskii equation
\begin{align}
 \left(i\partial_0+\muB\right)\Psi(t,\bm r)
 =-\frac{\bm\partial^2\Psi(t,\bm r)}{4m}
 +\gB\left|\Psi(t,\bm r)\right|^2\Psi(t,\bm r)\,. \label{eq:GP}
\end{align}
In order to consider a single vortex in the molecular superfluid, we set 
$\Psi(t,\bm r)=\sqrt{\nB}h(r)e^{i\lv\theta}$ in cylindrical
coordinates.
$\nB=n/2=\muB/\gB$ is the density of molecules far away from the vortex
core and $\lv$ is a winding number of the vortex.
$h(r)$ is some function of the radius $r$ from the vortex core, which
satisfies the boundary conditions
\begin{align}
 h(r\to0)\to0\qquad\text{and}\qquad h(r\to\infty)\to1\,. 
\label{eq:boundary}
\end{align}
Substituting $\Psi(t,\bm r)$ into Eq.~(\ref{eq:GP}), we obtain an
equation for $h(x)$, 
\begin{align}
 \frac{\partial^2h}{\partial x^2}+\frac1x\frac{\partial h}{\partial x}
 -\lv^2\frac{h}{x^2}+h-h^3=0\,, \label{eq:h(x)}
\end{align}
where $x=r/\xi$ is the dimensionless radius normalized by the healing
length of the molecular superfluid $\xi=1/\sqrt{4m\gB\nB}$. 

This equation should be solved under the boundary conditions
(\ref{eq:boundary}). 
Asymptotic forms of $h(x)$ at $x\to0$ and $x\to\infty$ are easily read
from Eq.~(\ref{eq:h(x)}) as follows:
\begin{align}
 h(x)&\propto x^{\lv}&&\text{for \ }x\to0\,,\\
 h(x)&=1-\frac{\lv^2}{2x^2}&&\text{for \ }x\to\infty\,. \label{eq:infinity}
\end{align}
The solutions of Eq.~(\ref{eq:h(x)}) connecting these two limits can be
obtained numerically, which are shown in Fig.~\ref{fig:GP-solution} for
$\lv=1,2$, and $3$. 

Because of the interaction between the molecule and the fermion, the
vortex structure in the molecular superfluid acts as a potential for
the fermion. 
If the interaction between the molecule and the fermion is repulsive
$(\gBF>0)$, the fermion will be attracted to the vortex core because
there are less molecules. 
Due to this effective attraction, bound fermion states on the vortex
turn out to be possible. 
The equation of motion from Eq.~(\ref{eq:fermion}) with the solution of 
Eq.~(\ref{eq:GP}) gives the Schr\"odinger equation for the fermion under
the vortex potential
\begin{align}
 \tilde E\psi(\bm r)=-\frac{\bm\partial^2\psi(\bm r)}{2m}
 +\gBF\nB\left\{h(r)^2-1\right\}\psi(\bm r)\,. \label{eq:BEC_limit}
\end{align}
$\tilde E$ is the energy eigenvalue of the fermion and is negative for
bound states.  

\begin{table}[tp]
 \caption{Bound energy levels $\epsilon_n$ and their ratios
 $\epsilon_n/\epsilon_{n-1}$ in the $S$-wave channel $(l=0)$ for
 $\gamma=1.47$. \label{tab:S-wave}}\smallskip 
 \begin{ruledtabular}
  \begin{tabular}{ccc}
   $\qquad n$&$\epsilon_n$&$\epsilon_n/\epsilon_{n-1}\quad$\\\hline
   $\qquad 0$&$-4.93\times10^{-1\phantom{0}}$&---$\quad\ $\\
   $\qquad 1$&$-3.39\times10^{-3\phantom{0}}$&$6.88\times10^{-3}\quad$\\
   $\qquad 2$&$-1.91\times10^{-5\phantom{0}}$&$5.64\times10^{-3}\quad$\\
   $\qquad 3$&$-1.08\times10^{-7\phantom{0}}$&$5.65\times10^{-3}\quad$\\
   $\qquad 4$&$-6.10\times10^{-10}$&$5.65\times10^{-3}\quad$\\
   $\qquad 5$&$-3.45\times10^{-12}$&$5.65\times10^{-3}\quad$\\
   $\qquad 6$&$-1.95\times10^{-14}$&$5.65\times10^{-3}\quad$\smallskip\\
   $\qquad\infty$&---&$5.65\times10^{-3}\quad$\\
  \end{tabular}
 \end{ruledtabular}
\end{table}

Separation of variables and use of the normalized radius $x=r/\xi$ lead
to the equation for the radial wave function $R(x)$ as follows: 
\begin{align}
 \epsilon R(x)=\left[-\frac{\partial^2}{\partial x^2}
 -\frac1x\frac{\partial}{\partial x}+\frac{l^2}{x^2}
 +\gamma\left\{h(x)^2-1\right\}\right]R(x)\,. \label{eq:radial}
\end{align}
Here $l$ is an angular momentum of the fermion and we have defined a
dimensionless energy eigenvalue $\epsilon=(2m\tilde E-k_z^2)\,\xi^2<0$
and a coupling ratio $\gamma=\gBF/2\gB$. 
$k_z$ is a momentum of the fermion along the vortex line. 
Note that Eq.~(\ref{eq:radial}) is invariant under $l\to-l$.  
This invariance cannot be exact because the vortex breaks parity.  
It is only an approximate property of the deep BEC limit, in which the
fermion field feels only the magnitude of the condensate but not its
phase [Eq.~(\ref{eq:BEC_limit})].

\subsubsection{Bound fermion states on a superfluid vortex}
The Schr\"odinger equation (\ref{eq:radial}) with the asymptotic
behavior of the vortex potential (\ref{eq:infinity}) predicts an
infinity number of energy levels for bound fermion states on a
superfluid vortex at least in the $S$-wave channel.
We can rewrite Eq.~(\ref{eq:radial}) far away from the vortex core
$(x\gg1)$ as  
\begin{align}
 \frac{\partial^2R}{\partial x^2}+\frac1x\frac{\partial R}{\partial x}
 -\left[\left|\epsilon\right|-\frac{\kappa^2}{x^2}\right]R=0
\end{align}
with $\kappa=\sqrt{\gamma\lv^2-l^2}$. 
This equation has the same form as Eq.~(\ref{eq:bessel}) so that the
discussion in Sec.~\ref{sec:BS} is applicable here. 
Consequently, the criterion for the angular momentum $l$ of the fermion 
in which bound fermion states are available is given by
\begin{align}
 \kappa^2=\gamma\lv^2-l^2>0\,.
\end{align}
An infinite number of energy levels for the bound fermion states appear
with each integer $l$ satisfying $l^2<\gamma\lv^2$ and their ratios
are asymptotically given by 
\begin{align}
 \frac{\epsilon_n}{\epsilon_{n-1}}=e^{-2\pi/\kappa}
 \qquad\quad\text{for \ }n\to\infty\,. \label{eq:ratio}
\end{align}
Whether we have bound fermion states with a certain $l$ depends only on
the value of the coupling ratio $\gamma=\gBF/2\gB=3\aBF/4\aB>0$. 
In particular, one can conclude that they are always possible for the
$S$-wave channel $(l=0)$ in the BEC limit. 

\begin{table}[tp]
 \caption{Bound energy levels $\epsilon_n$ and their ratios
 $\epsilon_n/\epsilon_{n-1}$ in the $P$-wave channel $(l=\pm1)$ for
 $\gamma=1.47$. \label{tab:P-wave}}\smallskip 
 \begin{ruledtabular}
  \begin{tabular}{ccc}
   $\qquad n$&$\epsilon_n$&$\epsilon_n/\epsilon_{n-1}\quad$\\\hline
   $\qquad 0$&$-1.57\times10^{-2\phantom{0}}$&---$\quad\ $\\
   $\qquad 1$&$-1.55\times10^{-6\phantom{0}}$&$9.86\times10^{-5}\quad$\\
   $\qquad 2$&$-1.68\times10^{-10}$&$1.09\times10^{-4}\quad$\\
   $\qquad 3$&$-1.83\times10^{-14}$&$1.09\times10^{-4}\quad$\smallskip\\
   $\qquad\infty$&---&$1.09\times10^{-4}\quad$\\
  \end{tabular}
 \end{ruledtabular}
\end{table}

Now we restrict ourselves to the energetically stable vortex $\lv=1$. 
Since the system is weakly coupled in the BEC limit $na^3\to0$, 
the scattering length between a molecule and a molecule or fermion is
calculable as a function of the fermion scattering length $a$ and
results in $\aB=0.60$--$0.75\,a$ \cite{pieri00,petrov03,ohashi05} or
$\aBF=1.179\,a$ \cite{skorniakov57,bedaque98,bulgac03}. 
These values leads to the coupling ratio $\gamma=1.18$--$1.47$, 
which predicts bound fermion states in the $P$-wave channel $(l=\pm1)$ as
well as in the $S$-wave channel. 
Hereafter we adopt the value $\aB=0.60\,a$ which has been confirmed by
a quantum Monte Carlo simulation \cite{astrakharchik04} and experiments
\cite{bartenstein04}. 
Then, the asymptotic ratios of bound energy levels
$\epsilon_\infty/\epsilon_{\infty-1}$ are given by $5.65\times10^{-3}$
for the $S$-wave channel and $1.09\times10^{-4}$ for the $P$-wave channel. 

\begin{figure*}[tp]
 \includegraphics[width=0.98\textwidth,clip]{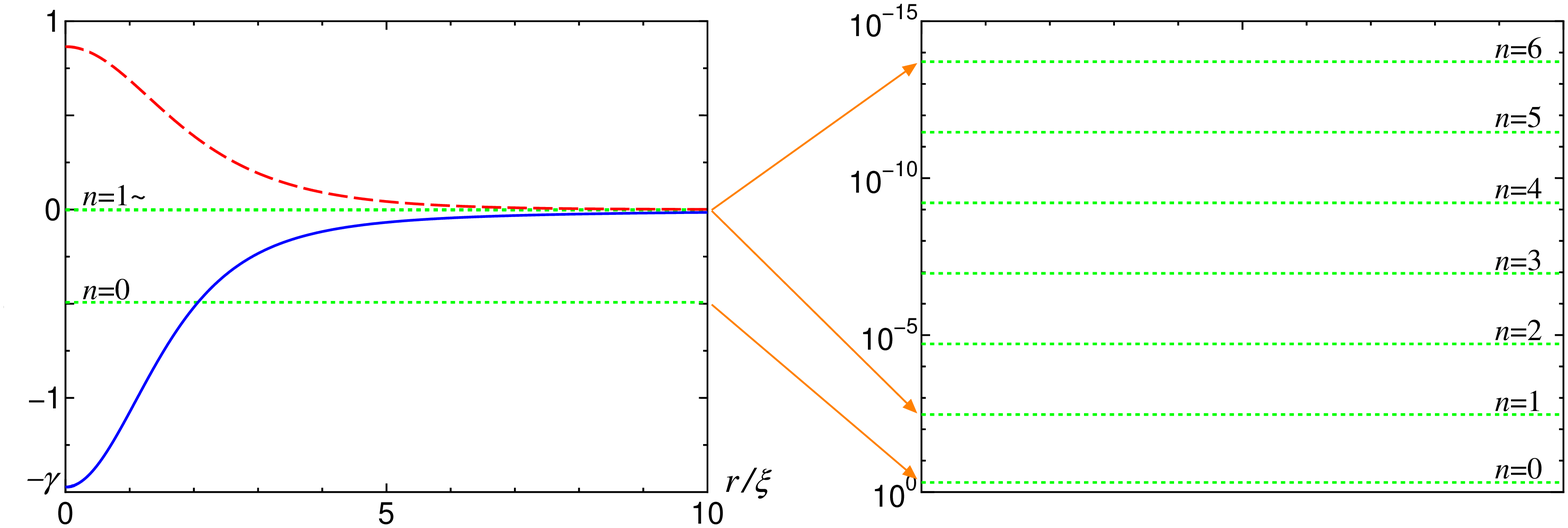}
 \caption{(Color online) Solutions of Eq.~(\ref{eq:radial}) in the
 $S$-wave channel $(l=0)$ for $\gamma=1.47$. 
 Left panel: Bound energy levels $\epsilon_n$ for $n=0,1,\dots,6$
 (dotted lines) and the potential energy $\gamma\left\{h(x)^2-1\right\}$ 
 (solid curve) are shown as functions of $x=r/\xi$. 
 The normalized wave function of the ground state $(n=0)$ is also shown
 by the dashed curve. 
 Right panel: $\left|\epsilon_n\right|$ (dotted lines) 
 are shown in the log scale. \label{fig:S-wave}}
\end{figure*}

\begin{figure*}[tp]
 \includegraphics[width=0.98\textwidth,clip]{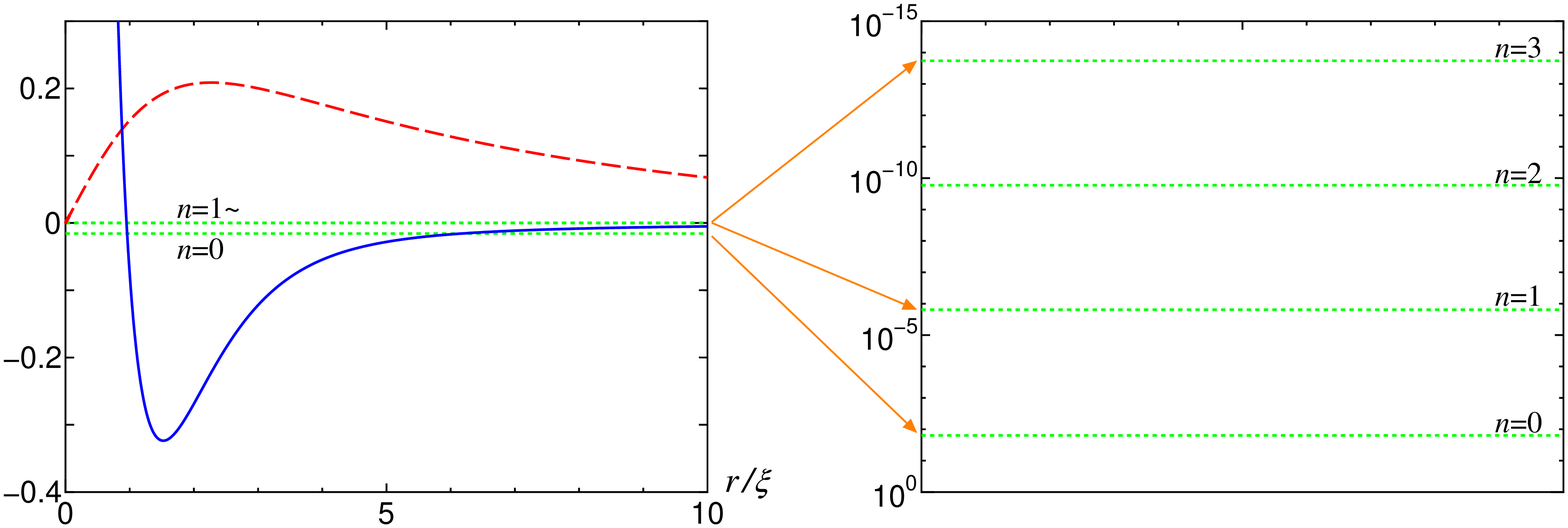} 
 \caption{(Color online) Solutions of Eq.~(\ref{eq:radial}) in the
 $P$-wave channel $(l=\pm1)$ for $\gamma=1.47$. 
 Left panel: Bound energy levels $\epsilon_n$ for $n=0,1,2,3$ (dotted 
 lines) and the potential energy $1/x^2+\gamma\left\{h(x)^2-1\right\}$
 (solid curve) are shown as functions of $x=r/\xi$. 
 The normalized wave function of the ground state $(n=0)$ is also shown
 by the dashed curve. 
 Right panel: $\left|\epsilon_n\right|$ (dotted lines) 
 are shown in the log scale. \label{fig:P-wave}}
\end{figure*}

The absolute values of binding energies should be determined by solving 
the Schr\"odinger equation (\ref{eq:radial}) numerically. 
Results on bound energy levels for $\epsilon_n\leq10^{-15}$ and their
ratios $\epsilon_n/\epsilon_{n-1}$ are listed in Table~\ref{tab:S-wave}
for the $S$-wave channel and in Table~\ref{tab:P-wave} for the $P$-wave
channel.  
The ratios of bound energy levels $\epsilon_n/\epsilon_{n-1}$ rapidly
converge to their asymptotic values from $n\agt2$ for both the $S$- 
and $P$-wave channels. 
Even in the first ratios $\epsilon_1/\epsilon_0$, their deviations from 
the asymptotic values are only 22\% for the $S$-wave channel and 9.1\% for 
the $P$-wave channel. 

The bound energy levels (dotted lines) are shown in Fig.~\ref{fig:S-wave}
for the $S$-wave channel and in Fig.~\ref{fig:P-wave} for the $P$-wave
channel as well as the potential energies
$l^2/x^2+\gamma\left\{h(x)^2-1\right\}$ (solid curves) as functions of
$x=r/\xi$. 
While the binding energies of excited states $(n\geq1)$ seems
degenerated into zero in the linear scale (left panels) for each
channels, they are allocated with equal intervals $2\pi/(\kappa\ln10)$
in the log scale (right panels) as indicated by Eq.~(\ref{eq:ratio}). 
The wave functions of the ground state $(n=0)$ normalized as
\begin{align}
 \int_0^\infty dx\,xR(x)^2=1
\end{align}
are also shown in each figure (dashed curves). 
In the $S$-wave channel, the ground state wave function is well localized
around the vortex core $r/\xi\alt5$ and its binding energy is comparable
to the bottom of potential energy
$\left|\epsilon_0\right|/\gamma=0.334$. 
Thus, we conclude that those bound fermion states in the BEC limit could
be measurable by future experiments at least in the ground state of the
$S$-wave channel. 

Away from the BEC limit, the fermion mass changes to the effective mass
$m^*$, which is larger than the bare mass $m^*>m$ \cite{son05}. 
Accordingly, the coupling ratio in the Schr\"odinger equation
(\ref{eq:radial}) changes to the effective one $\gamma^*=(m^*/m)\gamma$.
Because $\gamma^*>\gamma$, bound fermion states with higher angular
momenta are expected to appear as one departs from the BEC limit. 
However, if one is away from the BEC limit $na^3\agt1$, molecules
overlap each other and then the microscopic description with local
interactions Eqs.~(\ref{eq:boson}) and (\ref{eq:fermion}) will no
longer be valid. 
In order to treat the problem away from the BEC limit, the effective
field theory in terms of low-energy excitations is efficient. 

\subsubsection{Connection to effective field theory}
Let us see how the effective field theory in terms of low-energy
excitations, superfluid phonons and extra fermions, emerges from the
microscopic Lagrangians (\ref{eq:boson}) and (\ref{eq:fermion}).
For that purpose, we parametrize the field of bound molecule as 
$\Psi(t,\bm r)=\sqrt{\nB(t,\bm r)}e^{2i\varphi(t,\bm r)}$ and assume
that the magnitude $\nB(t,\bm r)$ is slowly varying in time and space,
which is satisfied far away from the vortex core. 
Then, Eqs.~(\ref{eq:boson}) and (\ref{eq:fermion}) can be rewritten by 
\begin{align}
 \LB+\LF&=\nB\biggl[\muB-2\dot\varphi
 -\frac{\left(\bm\partial\varphi\right)^2}m
 -\gBF\psi^\dagger\psi\biggr]-\frac{\gB}2\nB^2\nonumber\\
 &\quad+\frac{i\psi^\dagger\tensor\partial_0\psi}2
 -\frac{\left|\bm\partial\psi\right|^2}{2m}
 +\left(\mu+\Dmu\right)\psi^\dagger\psi\,. \label{eq:boson+fermion}
\end{align}
Since now the density of bound molecules $\nB$ is a variational
parameter, it should be determined by minimizing the action and results
in 
\begin{align}
 \gB\nB=\muB-2\dot\varphi
 -\frac{\left(\bm\partial\varphi\right)^2}m
 -\gBF\psi^\dagger\psi\,.
\end{align}

Substitution of this expression into Eq.~(\ref{eq:boson+fermion}) leads
to an effective theory in terms of the superfluid phonon $\varphi$ and
the fermion excitation $\psi$ as follows: 
\begin{align}
 \LB+\LF&\simeq P(\mu)
 +\frac{f_t^2}2\dot\varphi^2
 -\frac{f^2}2\left(\bm\partial\varphi\right)^2\nonumber\\
 &\quad+\frac{i\psi^\dagger\tensor\partial_0\psi}2
 -\frac{\left|\bm\partial\psi\right|^2}{2m}
 +\left(\mu+\Dmu\right)\psi^\dagger\psi \label{eq:matching}\\
 &\quad-\Delta(\mu)\,\psi^\dagger\psi+\frac{\partial\Delta}{\partial\mu}
 \biggl[\dot\varphi+\frac{\left(\bm\partial\varphi\right)^2}{2m}\biggr]
 \psi^\dagger\psi\,.\quad\nonumber
\end{align}
We have dropped a total derivative term and higher order terms in
$\varphi$. 
Here $P(\mu)=\muB^2/2\gB$ is identical to the pressure of the molecular 
superfluid up to an irrelevant constant, and $\Delta(\mu)=\gBF\muB/\gB$ 
is the energy cost to introduce a single fermion into the superfluid
originating in its interaction with the bound molecules. 
The low-energy parameters in the Lagrangian turn out to be given by 
\begin{align}
 f_t^2=\frac{\partial n}{\partial\mu}\,,\qquad f^2=\frac{n}{m}
 \qquad\text{with}\quad n=\frac{\partial P}{\partial\mu}\,.
\end{align}
This is the manifestation of general properties of Sec.~\ref{sec:EFT}
in the molecular superfluid case. 
The coupling between the superfluid phonons and extra fermions is also
given by the derivative of the fermion energy cost
$\Delta'(\mu)=2\gBF/\gB$ as in Eq.~(\ref{eq:Lgap}).

\subsection{Away from the BEC limit}

\subsubsection{Effective field theory}
As mentioned above, the problem of writing down an effective field
theory that couples the superfluid phonons $\varphi$ and the extra
fermions $\psi$ is identical to the same problem in the $^3$He-$^4$He
mixture.   
This is because their symmetries and the pattern of symmetry breaking in
the two cases are the same. 
Let us elaborate on this point in more detail.

In the $^3$He-$^4$He mixed system, there are two conserved charges: the 
number of $^4$He atoms $N_4$ and the number of $^3$He atoms $N_3$.
The first charge is spontaneously broken by the $^4$He superfluid
ground state.  The order parameter carries a unit $N_4$ charge, but is
neutral with respect to the $N_3$ charge.  
The $^3$He atoms carry only the $N_3$ charge which is unbroken, while 
being neutral with respect to the $^4$He charge.

In the slightly polarized fermion system, the number of spin-up
fermions $N_\uparrow$ and the number of spin-down fermions
$N_\downarrow$ are conserved separately.  We suppose all extra fermions 
have spin up.  It is, however, more convenient to use another basis,
consisting of $N_\downarrow$ and the total polarization
$Y=N_\uparrow-N_\downarrow$.  
The Cooper pairs carry unit $N_\downarrow$ charge and its condensation
spontaneously breaks this symmetry. 
The extra fermions carry a unit $Y$ charge but are neutral with respect
to $N_\downarrow$.  
Thus, $N_\downarrow$ is equivalent to $N_4$ in the $^3$He-$^4$He
mixture, and $Y=N_\uparrow-N_\downarrow$ is equivalent to $N_3$.

It is now possible to write down the effective Lagrangian for the case
at hand by direct analogy with the case for $^3$He-$^4$He mixture.  
We simply need to make the following replacement for the particle masses
and chemical potentials:
\begin{alignat}{2}
 m_4&\rightarrow 2m,\qquad&m_3&\rightarrow m, \\
 \mu_4&\rightarrow 2\mu,\qquad&\mu_3&\rightarrow\mu+H. 
 \label{eq:mu43muH}
\end{alignat}
The equations in Eq.~(\ref{eq:mu43muH}) follow from the requirement that 
$\mu_4n_4+\mu_3n_3\to\mu_\uparrow n_\uparrow+\mu_\downarrow n_\downarrow$
when $n_4\to n_\downarrow$ and $n_3\to n_\uparrow-n_\downarrow$.
For convenience, we shall also make some rescalings so as to absorb
extra factors 2 appearing in the equations:
\begin{equation}
  \varphi \to 2\varphi,\qquad2\eta\to\eta. \label{eq:rescale}
\end{equation}
As a result, the Galilean invariant effective Lagrangian (\ref{eq:Leff})
becomes 
\begin{equation}
\begin{split}
 \Leff&=\frac{f_t^2}2\dot{\varphi}^2
 -\frac{f^2}2\left(\bm\partial\varphi\right)^2
 +\frac{i\psi^\dagger\tensor\partial_0\psi}2
 -\frac{\bigl|\bm\partial\psi\bigr|^2}{2m^*}\\
 &\quad+\left(\mu+\Dmu-\Delta\right)\psi^\dagger\psi 
 \label{eq:effective}\\
 &\quad+g_1\bm\partial\varphi\cdot
 \frac{i\psi^\dagger\tensor{\bm\partial}\psi}{2m}
 +\biggl[g_2\dot\varphi+g_3\frac{\left(\bm\partial\varphi\right)^2}
 {2m}\biggr]\psi^\dagger\psi\,.
\end{split}
\end{equation}
Here $m^*$ is the effective fermion mass and the couplings in the
effective Lagrangian are given by 
\begin{align}
 g_1&=\eta\frac{m}{m^*}\,,&
 g_2&=\frac{\partial\Delta}{\partial\mu}-\eta\,,&
 g_3&=\frac{\partial\Delta}{\partial\mu}-\eta^2\frac{m}{m^*}\,,
 \label{eq:coupling}
\end{align}
with $\eta$ being defined by $m^*=(1+\eta)m$. 
The universal relation among these couplings
\begin{align}
  g_3=g_2+g_1
\end{align}
is a consequence of the Galilean invariance. 
In particular, $g_1=0$ and $g_2=g_3=\partial\Delta/\partial\mu=4\gamma$ 
in the BEC limit $m^*\to m$, which reproduce the result of
Eq.~(\ref{eq:matching}). 

\subsubsection{Bound fermion states on a superfluid vortex}
Let us consider the bound fermion states on a superfluid vortex with
its winding number $\lv$. 
The discussion is parallel to that in Sec.~\ref{sec:BS}. 
What we have to be careful here is that the phase $\varphi$ of the
condensate around the vortex is given by $\varphi(t,\bm r)=\lv\theta/2$
in cylindrical coordinates. 
Extra one-half compared to the $^3$He-$^4$He mixture case is because 
$\varphi$ is normalized to be half of the phase of the Cooper
pair $\Psi=|\Psi|e^{2i\varphi}$ in Eq.~(\ref{eq:rescale}) for the
fermion superfluid case. 
The equation of motion for the fermion field $\psi$ from
Eq.~(\ref{eq:effective}) gives a Schr\"odinger equation as follows:
\begin{align}
 \tilde E\psi(\bm r)=-\frac{\bm\partial^2\psi(\bm r)}{2m^*}
 -\frac{ig_1\lv}{2mr^2}\frac{\partial\psi(\bm r)}{\partial\theta}
 -\frac{g_3\lv^2}{8mr^2}\psi(\bm r)\,.
\end{align}
The third term corresponds to the $1/r^2$ vortex potential far away from 
the vortex core in Eq.~(\ref{eq:BEC_limit}) in the BEC limit. 
On the other hand, the second term, which was absent in the BEC limit,
is required by the Galilean invariance when $m^*\neq m$.
This term represents that the fermion with angular momentum in the
opposite direction to the vortex's one is energetically favored, because
it decrease the total angular momentum of the system. 

Separation of variables leads to the equation for the radial wave
function $R(r)$ as follows: 
\begin{align}
 &\left(2m^*\tilde E-k_z^2\right)R(r)\\
 &\quad=\left[-\frac{\partial^2}{\partial r^2}
 -\frac1r\frac{\partial}{\partial r}
 +\frac{l^2}{r^2}+g_1\frac{m^*}m\frac{\lv l}{r^2}
 -g_3\frac{m^*}{4m}\frac{\lv^2}{r^2}\right]R(r)\,. \nonumber
\end{align}
Here $l$ is an angular momentum of the fermion and $k_z$ is its momentum
along the vortex line. 
As we have proved in Sec.~\ref{sec:BS}, whether we have bound fermion
states for a certain $l$ is determined only by the sign of the
coefficient of $1/r^2$. 
They are possible for angular momenta $l$ which satisfy the criterion
\begin{align}
 \kappa^2=g_3\frac{m^*}{4m}\lv^2-g_1\frac{m^*}{m}\lv l-l^2>0\,.
 \label{eq:criterion}
\end{align}
Using the definitions of the couplings (\ref{eq:coupling}) and $\eta$,
$m^*=(1+\eta)m$, the criterion can be rewritten as $l_-<l<l_+$ with 
\begin{align}
 \frac{l_\pm}{\lv}=\frac12-\frac{m^*}{2m}\pm\frac12\sqrt{\frac
 {\partial\Delta}{\partial\mu}\frac{m^*}{m}}\,. \label{eq:angular_mom}
\end{align}
An infinite number of bound energy levels with their asymptotic ratio
$e^{-2\pi/\kappa}$ appear for each integer $l$ satisfying $l_-<l<l_+$. 

\subsubsection{Discussion on parameters and conjectures}
We have two free parameters which cannot be determined within our
effective field theory, the effective fermion mass $m^*/m$ and the
derivative of fermion energy cost $\Delta'(\mu)$ in the superfluid. 
These two parameters are functions of the fermion scattering length $a$
and the fermion chemical potential $\mu$ or density $n$. 
They should be measurable by experiments in the same way as $m_3^*$ and 
$\alpha_0=\Delta'(\mu_4)-1$ in the $^3$He-$^4$He mixtures.
While such measurements are not achieved yet, we still know some general
properties of those quantities. 

As we have discussed, the effective fermion mass $m^*(\mu,a)$
coincides with its bare mass $m$ in the BEC limit $na^3\ll1$.  On the
other hand, it has been argued that $m^*(\mu,a)$ becomes infinite at
the so-called splitting point (SP), located on the BEC half ($a>0$) of
the crossover diagram~\cite{son05}.  Therefore, it is natural to
assume that $m^*(\mu,a)/m$ is an increasing function of $na^3$ from
unity (the BEC limit) to infinity (the SP limit), as $na^3$ increases
from zero to some critical value.

As for the derivative of fermion energy cost $\Delta'(\mu,a)$, it is
presumably positive in the BEC regime. 
This can be seen by writing the derivative of fermion energy cost as
\begin{align}
 \frac{\partial\Delta(\mu,a)}{\partial\mu}=\frac{\partial n}{\partial\mu}
 \frac{\partial\Delta(n,a)}{\partial n}\,.
\end{align}
Thermodynamic stability implies $\partial n/\partial\mu>0$. 
The energy cost $\Delta$ to introduce a fermion into the superfluid
originates in the interaction of the fermion with bound molecules in the
BEC regime. 
Their interaction is considered to be repulsive because the Pauli
principle between the extra fermion and a fermion in the bound molecule
with the same sign of spin acts as an effective repulsion. 
Thus, the fermion energy cost will be positive in the superfluid and an
increasing function of the density $n$, which results in
$\partial\Delta/\partial n>0$. 
In fact, this is the case in the BEC limit where $\Delta=\gBF\nB$ with 
the repulsive coupling $\gBF=1.179\,a>0$.  
It is also true at the unitarity limit ($na^3=\infty$) where 
$\Delta\sim n^{2/3}$.
Accordingly, we conclude that $\Delta'(\mu)$ is positive in the BEC
regime. 

With the use of those properties on the parameters, we can derive some
interesting predictions from the formula (\ref{eq:angular_mom}). 
Let us restrict ourselves to the energetically stable vortex $\lv=1$. 
Because of $l_\pm=\sqrt{\gamma}=1.21$ in the BEC limit, the $S$-wave
$(l=0)$ and $P$-wave $(l=\pm1)$ bound fermion states are possible. 
On the other hand, $l_\pm$ goes to negative infinity in the SP limit
$l_\pm\to-\infty$, because $m^*/m\to\infty$ and
$\partial\Delta/\partial\mu$ should be finite without phase transitions. 
Therefore, $l_-$ must get across integers less than $-1$ where new bound 
energy levels appear, and $l_+$ must get across integers not greater
than 1 where existing bound energy levels disappear between the BEC and
SP limits. 
In particular, an infinite number of negative and large angular momenta 
become ready for the bound fermion states in the vicinity of the SP
limit. 

\begin{figure}[tp]
 \includegraphics[width=0.45\textwidth,clip]{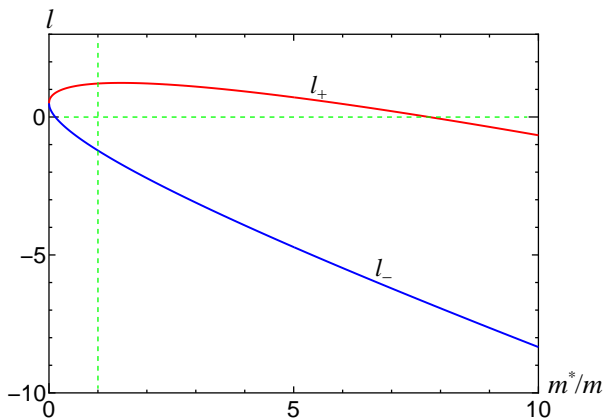}
 \caption{(Color online) $l_+$ and $l_-$ in Eq.~(\ref{eq:angular_mom})
 as functions of $m^*/m$ for $\partial\Delta/\partial\mu=4\gamma$.
 $m^*/m=1$ corresponds to the BEC limit, while $m^*/m\to\infty$
 corresponds to the splitting point in Ref.~\cite{son05}. 
 The curves are shown for $m^*/m\geq0$. \label{fig:angular_mom}}
\end{figure}

For illustration, let us employ a somewhat unjustified ansatz,
$\Delta'(\mu,a)=4\gamma$, which is an extrapolation from the
weak coupling BEC limit.
Using the value $\gamma=1.47$, $l_\pm$ in Eq.~(\ref{eq:angular_mom})
are evaluated as functions of $m^*/m$ in Fig.~\ref{fig:angular_mom}. 
Bound fermion states are possible with angular momenta $l$ which are
integers between $l_-$ and $l_+$.  
The term proportional to $g_1$ in Eq.~(\ref{eq:effective}) suppresses
bound fermions with positive angular momenta, while it enhances bound
fermions which have negative angular momenta in the opposite direction
to the vortex's one. 
Fig.~\ref{fig:angular_mom} shows that bound fermion states with positive 
angular momentum disappear away from the BEC limit, while those with
arbitrary higher negative angular momenta become possible as one
approaches to the SP limit $m^*/m\to\infty$. 

The fermion effective masses where bound fermion states appear
$m^*/m\,|_{l_-=l}$ or disappear $m^*/m\,|_{l_+=l}$ are estimated for
the first four angular momenta $l$ in Table~\ref{tab:estimation}.
Especially, the $S$-wave bound fermion states which have the deepest
binding energy in the BEC limit are found to disappear when $g_3$
changes its sign at $m^*/m=7.77$.  We emphasize that quantitative
results in Fig.~\ref{fig:angular_mom} and Table~\ref{tab:estimation} may
not be reliable due to the ansatz we employed.

Finally, it would be interesting to compare our results in the polarized 
Fermi gas with those in the unpolarized Fermi gas from the 
Bogoliubov--de~Gennes approach \cite{machida05,sensarma05}. 
The results in Ref.~\cite{sensarma05} show that bound fermion states on
a vortex are possible only for $l=0,-1$ at $na^3=1/3\pi^2$ in the BEC 
regime, while bound fermion states become possible for all $l\leq0$ in
the unitarity limit $na^3=\infty$. 
The tendency that possible angular momenta for bound states increase
as one gets away from the BEC limit is consistent with what we found, 
however, we predict $l=1$ bound fermion states as well as $l=0,-1$ in
the BEC limit. 
How these two results match with each other as a function of
the polarization will be an interesting future problem. 

\begin{table}[tp]
 \caption{Estimated values of the fermion effective mass where bound
 fermion states appear $m^*/m\,|_{l_-=l}$ and disappear
 $m^*/m\,|_{l_+=l}$ for several angular momenta $l$. 
 \label{tab:estimation}}\smallskip 
 \begin{ruledtabular}
  \begin{tabular}{cccc}
   $\quad l$&$\left.m^*/m\,\right|_{l_-=l}\quad$
   &$\quad l$&$\left.m^*/m\right|_{l_+=l}\quad$\\\hline
   $\quad-2$&1.77$\quad$&$\quad\phantom{-}1$&3.62$\quad$\\
   $\quad-3$&2.88$\quad$&$\quad\phantom{-}0$&7.77$\quad$\\
   $\quad-4$&4.09$\quad$&$\quad-1$&11.1$\quad$\\
   $\quad-5$&5.37$\quad$&$\quad-2$&17.0$\quad$
  \end{tabular}
 \end{ruledtabular}
\end{table}

\section{Summary and conclusions}
In this paper, we constructed low-energy effective field theories for
dilute fermion excitations in superfluids with emphasis on the Galilean
invariance [Eq.~(\ref{eq:Leff}) for boson superfluids or
Eq.~(\ref{eq:effective}) for molecular superfluids]. 
The Galilean invariance as well as the global symmetries of the system
considerably restrict the possible form of the effective Lagrangian. 
We showed three terms representing interactions between the fermion
excitations $\psi$ and superfluid phonons $\varphi$ appear to the lowest 
nontrivial order of the fields. 
The couplings for the three interaction terms are not independent as a
consequence of the Galilean invariance. 
They are written by the effective fermion mass $m^*/m$ and the
derivative of the fermion energy cost with the chemical potential
$\partial\Delta/\partial\mu$ [Eq.~(\ref{eq:g_123}) or
(\ref{eq:coupling})], which are measurable quantities by experiments. 

We consider that the effective Lagrangian obtained here is valid as long
as the following conditions are satisfied. (i) The low-energy dynamics
of superfluids is dominated by the excitations of superfluid phonons
or, in other words, the variations in the magnitude of the condensate are 
negligible. (ii) The fermion's dispersion is given by the quadratic power
of its momentum with a positive coefficient as in
Eq.~(\ref{eq:dispersion}). Higher order corrections to the dispersion
are negligible in the low-energy dynamics of fermions. 
(iii) Fermions are dilute or weakly coupled so that their
self-interactions are negligible. 

With the use of the effective field theory, we studied bound fermion
states on a single vortex of the superfluid.
We derived a simple criterion for an angular momentum $l$ of fermion in
which bound fermion states are available [Eq.~(\ref{eq:criterion_He}) or
(\ref{eq:criterion})]. 
An infinite number of bound energy levels appear for angular
momenta satisfying the criterion and their ratios are asymptotically
given by Eq.~(\ref{eq:ratio_He}). 

We applied our effective field theory to two boson-fermion mixed
systems, a dilute solution of $^3$He in $^4$He superfluid and 
the BEC regime of a cold polarized Fermi gas. 
For the $^3$He-$^4$He mixture, we determined parameters of the
effective theory from experimental data as functions of pressure. 
As a result, we predict that bound $^3$He states with $l=-2,-1,0$ 
will be realized on a vortex of the $^4$He superfluid in a whole range
of pressure, $0$--$20$ atm, where experimental data are available.
Asymptotic ratios of bound energy levels are calculated in
Table~\ref{tab:ratio} for each angular momentum.
Those properties should be in principle confirmed by future accurate
experiments. 

As for the cold polarized Fermi gas, we determined parameters of
the effective field theory from the microscopic description in the BEC 
limit. 
As a consequence, $S$-wave $(l=0)$ and $P$-wave $(l=\pm1)$ bound fermion
states turned out to be realized in the BEC limit. 
Since the fermion mass compared to its bare mass $m^*/m$ could change
from unity (the BEC limit) to infinity (the SP limit) in the BEC regime, 
bound fermion states with arbitrary higher negative angular momentum
will become available away from the BEC limit. 
Especially, the bound fermion states with $l=0,\pm1$, which are realized
in the BEC limit, are shown to disappear as one get away from the BEC
limit. 

While we have concentrated on the study of bound fermion states on a
superfluid vortex in this paper, our effective field theory will be
useful to study other problems in the boson-fermion mixtures. 
Also, our effective field theory is universal to any other boson-fermion
mixed systems. 
Those investigations should be performed in future works.

\begin{acknowledgments}
 Y.~N. is supported by the Japan Society for the Promotion of Science
 for Young Scientists.  This work is supported, in part, by DOE Grant 
 No.~DE-FG02-00ER41132. 
\end{acknowledgments}

\appendix*
\section{The Bardeen-Baym-Pines parameter}
Here let us remind of the relation between the energy cost $\Delta$ to
introduce a $^3$He quasiparticle into the $^4$He superfluid in
Eq.~(\ref{eq:Lgap}) and the relative fractional molar volume of a dilute 
solution of $^3$He in $^4$He, or the Bardeen-Baym-Pines (BBP)
parameter, $\alpha$ \cite{bardeen67,baym-pethick}. 
$\alpha$ is a directly measurable quantity by experiments. 
It is sufficient for the present purpose to limit ourselves at zero
temperature for simplicity. 

BBP parameter $\alpha(x,P)$ at the $^3$He concentration $x$ and the
pressure $P$ is defined through the following equation:
\begin{align}
 V_{34}(x,P)=V_4(P)\left[1+x\alpha(x,P)\right], \label{app:V34}
\end{align}
where $V_{34}$ and $V_4$ are the molar volume of the dilute solution of 
$^3$He in $^4$He and the pure $^4$He, respectively. 
Total number of $^3$He and $^4$He atoms $N$ is fixed here. 
Since we have interest in the dilute limit of $\alpha$, let us rewrite
Eq.~(\ref{app:V34}) as
\begin{align}
 \alpha_0(P)\equiv\lim_{x\to0}\frac{V_{34}(x,P)-V_4(P)}{x\,V_4(P)}\,,
\end{align}
and calculate $V_{34}$ in the leading order of $x$. 

The chemical potential of the mixture is defined as
\begin{align}
 \mu(x,P)=x\mu_3(x,P)+(1-x)\mu_4(x,P) \label{app:mu_old}
\end{align}
with $\mu_3$ and $\mu_4$ being the chemical potentials of the $^3$He and
$^4$He atoms. 
We can calculate the molar volume of the mixture by differentiating the
mixture chemical potential with the pressure as
\begin{align}
 V_{34}(x,P)=A\!\left.\frac{\partial\mu}{\partial P}\,\right|_x\,,
\end{align}
where $A$ is the Avogadro's constant. 
With the use of the Gibbs--Duhem relation
\begin{align}
 x\left.\frac{\partial\mu_3}{\partial x}\,\right|_P
 +(1-x)\left.\frac{\partial\mu_4}{\partial x}\,\right|_P=0\,,
\end{align}
Eq.~(\ref{app:mu_old}) is rewritten as
\begin{align}
 \mu(x,P)&=(1-x)\mu_4(0,P)+x\mu_3(x,P)\nonumber\\
 &\quad-(1-x)\int_0^x dx'\frac{x'}{1-x'}
 \frac{\partial\mu_3(x',P)}{\partial x'}\,.  \label{app:mu_new}
\end{align}

The energy of the $^3$He quasiparticle excitation $E(\bm k)$ can be 
expressed with its effective mass $m_3^*(P)$ as 
\begin{align}
 E(\bm k)=\Delta(P)+\frac{\bm k^2}{2m_3^*(P)}+xE_\mathrm{int}(x,P,\bm k)\,.
\end{align}
The first term $\Delta$ is the energy cost to introduce a single $^3$He
quasiparticle to the pure $^4$He superfluid, and the third term
proportional to $x$ represents the contribution of $^3$He-$^3$He
interaction to the excitation energy. 
Since the third term gives higher order corrections of the order of
$O(x^2)$ to $\alpha(x,P)$, we neglect it hereafter. 
Therefore, the $^3$He chemical potential in the mixture is of the form: 
\begin{align}
 \mu_3(x,P)=\Delta(P)+\mu_\mathrm{F}(x,P)\,. \label{app:mu3}
\end{align}
$\mu_\mathrm{F}$ is the chemical potential of a free Fermi gas of mass
$m_3^*$ and density $xN$, which is simply given, at zero temperature, by
\begin{align}
 \mu_\mathrm{F}(x,P)=\frac{(3\pi^2N)^{2/3}}{2m_3^*(P)}x^{2/3}
 =\mu_\mathrm{F}(1,P)x^{2/3}\,. \label{app:muF}
\end{align}

Substituting Eqs.~(\ref{app:mu3}) and (\ref{app:muF}) into
Eq.~(\ref{app:mu_new}), we obtain 
\begin{align}
 \mu(x,P)&=(1-x)\mu_4(0,P)+x\Delta(P)\nonumber\\
 &\quad+\frac35\,x^{5/3}\,\mu_\mathrm{F}(1,P)+O(x^2)\,.
\end{align}
The differentiation of this equation with $P$ leads to the
expression of the molar volume of the mixture as follows:
\begin{align}
  V_{34}(x,P)=V_4(P)-xV_4(P)+xA\frac{\partial\Delta(P)}{\partial P}
 +O(x^{5/3})\,,
\end{align}
where $V_4(P)=A\,\partial\mu_4(0,P)/\partial P$.
Consequently, BBP parameter at zero $^3$He concentration is given by 
\begin{align}
 \alpha_0(P)=\frac{A}{V_4(P)}\frac{\partial\Delta(P)}{\partial P}-1
 =\frac{\partial\Delta(P)}{\partial\mu_4(0,P)}-1\,. \label{app:alpha0}
\end{align}

$\alpha_0$ has been determined as a function of the pressure $P$ by the 
experiment \cite{hatakeyama03} up to about 10 atm, and found to be
fitted very well by the following polynomial
[Eq.~(60) in \cite{hatakeyama03}]: 
\begin{align}
 \alpha_0(P)=\sum_{i=0}^4 a_i P^i \label{app:BBP}
\end{align}
with
\begin{equation}
\begin{split}
 a_0&=\phantom{-}2.88069661\\
 a_1&=-1.26990768\times10^{-2}\quad\text{atm}^{-1}\\
 a_2&=\phantom{-}9.57212464\times10^{-4}\quad\text{atm}^{-2}\\
 a_3&=-6.13373794\times10^{-5}\quad\text{atm}^{-3}\\
 a_4&=\phantom{-}2.26173741\times10^{-6}\quad\text{atm}^{-4}\,. 
\end{split}
\end{equation}
The unit is converted from ``kgf/cm$^2$'' in \cite{hatakeyama03} to
``atm'' here.
The values of $\alpha_0(P)$ for $P=0,5$ and $10$ atm are shown in
Table~\ref{tab:pressure}. 

Finally, we note that $\Delta$ and $\mu_4$ in Eq.~(\ref{app:alpha0}) are
identical to $\Delta$ and $\mu_4$ introduced in Eq.~(\ref{eq:Lgap}) of
the text. 
Therefore, the coupling between the $^3$He quasiparticle and the $^4$He
superfluid phonons $\partial\Delta/\partial\mu$ is related with the BBP 
parameter at zero concentration $\alpha_0$ by
$\partial\Delta/\partial\mu=\alpha_0+1$.

\end{document}